\documentclass[graybox]{svmult}

\usepackage{type1cm}        
\usepackage{makeidx}         
\usepackage{graphicx}        
\usepackage{multicol}        
\usepackage[bottom]{footmisc}

\usepackage{newtxtext}       %
\usepackage{newtxmath}       

\makeindex             

\begin{document}

\title*{A brief introduction to the scaling limits and effective equations in kinetic theory}
\author{ M. Pulvirenti and S. Simonella }
\institute{ M. Pulvirenti   \at Dipartimento di Matematica, Universit\`a di Roma La Sapienza, 
Piazzale Aldo Moro 5, 00185 Roma, Italy; \\
 International Research Center M\&MOCS, Universit\`a dell'Aquila, Palazzo Caetani, 04012 Cisterna di Latina (LT), Italy. \\ \email{pulviren@mat.uniroma1.it}
\and S. Simonella \at ENS de Lyon, UMPA UMR 5669 CNRS, 46 all\'{e}e d'Italie,
69364 Lyon Cedex 07, France \\ \email{sergio.simonella@ens-lyon.fr}}
%
%
\maketitle

\abstract*{These lecture notes provide the material for a short 
introductory course on effective equations for classical particle systems. They concern the basic equations in kinetic theory, written by Boltzmann and Landau, describing rarefied gases and weakly interacting plasmas respectively. These equations can be derived formally, under suitable scaling limits, taking classical particle systems as a starting point. A rigorous proof of this limiting procedure is difficult and still largely open. We discuss some mathematical problems arising in this context.}

\abstract {These lecture notes provide the material for a short 
introductory course on effective equations for classical particle systems. They concern the basic equations in kinetic theory, written by Boltzmann and Landau, describing rarefied gases and weakly interacting plasmas respectively. These equations can be derived formally, under suitable scaling limits, taking classical particle systems as a starting point. A rigorous proof of this limiting procedure is difficult and still largely open. We discuss some mathematical problems arising in this context.}

\section{The foundations of kinetic theory}
\label{sec:1}

Many interesting systems in physics and applied sciences consist of a large number of identical components so that they are difficult to analyze from a mathematical point of view. On the other hand, quite often, we are not interested in a detailed description of the system but rather in its collective behaviour. Therefore, it is necessary to look for all procedures leading to simplified models, retaining the interesting features of the original system, cutting away redundant information. This is  the methodology of statistical mechanics and of kinetic theory. Here we want to outline the limiting procedure leading from the microscopic description of a large particle system (based on the fundamental laws like the Newton or Schr\"{o}dinger equations) to the more practical picture dictated by kinetic theory.

Although the methods of kinetic theory are frequently applied to a large variety of complex systems (consisting of a huge number of individuals), we will discuss only models arising in physics and more precisely in classical mechanics. The starting point is a system of $N$ identical particles in the physical space. A microscopic state of the system is a sequence $z_1,\cdots,z_N$ where $z_i = (x_i,v_i)$ denotes position and velocity of the $i$-th particle. The equations of motion are given by Newton's laws of dynamics.

We are interested in a situation where $N$ is very large (for instance, a cubic centimeter of a rarefied gas contains approximately $10^{19}$ molecules). The knowledge of the microscopic states becomes useless, and we turn to a statistical description. We introduce a probability measure $W^N_0( Z_N ) d Z_N$ (absolutely continuous with respect to the Lebesgue measure), defined on the phase space $\mathbb{R}^{3N}\times \mathbb{R}^{3N}$, where $$Z_N =(z_1,\cdots,z_N)=(x_i,v_i,\cdots,x_N,v_N)\;.$$
$W^N_0$ assigns the same statistical weight to two different vectors $Z_N$ and $Z'_N$ differing only for the order of particles, i.e., identifying the same physical configuration. 

The time-evolved measure is defined by
\begin{equation}\label {meas}
W^N(Z_N,t)=W^N_0( \Phi ^{-t }(Z_N) )\;.
\end{equation}
Here $ \Phi^t( Z_N )$   denotes the dynamical, measure-preserving flow constructed by solving the equations of motion.

We can establish a partial differential equation, called the Liouville equation, describing the evolution of the measure \eqref{meas}. However, this equation is also not tractable in practice. To have an efficient reduced description, one can focus on the time evolution for the probability distribution of a {\em given} particle (say particle 1), all the particles being identical.
To this end, we define the $j$-particle marginals
\begin{equation}\label{marg}
f^N_j (Z_j,t) := \int_{\mathbb{R}^{3N}\times \mathbb{R}^{3N}} dz_{j+1} \cdots dz_N W^N(Z_j,z_{j+1},\cdots, z_N,t)\;,\qquad j=1,\cdots,N\;,
\end{equation}
and we look for an equation describing the evolution of $f^N_1$.
We deduce, in most of the physically relevant situations, an evolution equation of the form
\begin{equation}\label{hie11}
\partial_t f_1^N  =-v\cdot \nabla f^N_1 + Q\;.
\end{equation}
The first term in the right-hand side is due to the free transport of particles, while the term $Q$ should describe the interaction of particle $1$ with the rest of the system.

We face a big difficulty. Since the interaction is binary, $Q$ will depend on $f_2^N$, namely the two-particle marginal. In other words, \eqref{hie11} is still useless: to know $f_1^N$ we need to know $f_2^N$, and to know $f_2^N$ we need to know $f_3^N$, and so on. We handle a hierarchy of equations, called BBGKY hierarchy 
(from the names of the physicists Bogolyubov, Born, Green, Kirkwood, Yvon).

Here enters the property called {\em propagation of chaos}, that is,
\begin{equation}
\label{PC1}
f^N_2(x_1,v_1, x_2, v_2, t) \simeq f^N_1(x_1,v_1, t) f^N_1( x_2, v_2, t). 
\end{equation}
Accepting \eqref{PC1}, $Q$ becomes an operator acting on $f_1^N$ and \eqref{hie11} is a closed equation. We have thus replaced a huge ordinary differential system by a single PDE. The price we pay is that \eqref{hie11} is nonlinear.

The equality in Eq.\,\eqref{PC1} is certainly false, since it expresses the statistical independence of particle $1$ and particle $2$ which, even if assumed at time $0$, cannot hold at later times. Indeed, the dynamics creates correlations. Nevertheless, one can hope to recover this property in some {\em asymptotic} situation described by a suitable scaling limit. This is what happens in two different physical contexts: the low-density and the weak-coupling limits, yielding two different kinetic equations, namely the Boltzmann and the Landau equations, respectively.
The passage from hamiltonian mechanics to this kinetic description is actually very delicate. As we shall see later on, we go from a deterministic time-reversible system to an irreversible equation. 

A different scaling procedure is the so-called mean-field limit. This leads to the Vlasov equation, which has still a time-reversible, hamiltonian nature. It is a sort of continuum limit and hence much simpler than the previous two. Some challenging and interesting problems concerning the mean-field limit are anyway still open, but we shall not discuss them in this note.

\section{Low-density limit and Boltzmann equation}
\label{sec:2}

Ludwig Boltzmann established an evolution equation to describe the behaviour of a rarefied gas in 1872, starting from the mathematical model of elastic balls and using mechanical and statistical considerations \cite{Bo64}.
The importance of this equation is twofold. On one side, it provides (as well as the hydrodynamical equations) a reduced description of the microscopic world. On the other, it is also an important tool for applications, especially for dilute fluids when the hydrodynamical equations fail to hold.

According to the general paradigm of kinetic theory, the starting point of Boltzmann's analysis is to renounce to study the gas in terms of the detailed motion of the molecules of the full system. It is preferable to investigate a function
$f(x,v)$, the probability density of a given particle, where $x$ and $v$ denote its position and velocity. 
Or, following the original approach proposed by Boltzmann,
$f(x,v) dx dv$ is to rather be interpreted as the fraction of molecules happening to be in the cell
of the phase space of size $dx dv$ around $(x,v)$. 
The two quantities are not
exactly the same, but they are asymptotically equivalent (when the number of
particles diverges) if a {\em law of large numbers} holds. 

Boltzmann considered a gas as microscopically described by a system of elastic (hard) balls,
colliding according to the laws of classical mechanics.
In this case, the  {\em Boltzmann equation} for the one-particle distribution function reads
\begin{equation}\label{beq}
(\partial_t+v\cdot \nabla_x)f=Q_B(f,f)
\end{equation}
where $Q_B$, the collision operator, is defined by 
\begin{equation}\label{bcoll}
Q_B(f,f)(x,v) := \int_{\mathbb{R}^3} dv_1 \int_{S_+^2} dn  \  (v-v_1)\cdot n \ 
[f (x,v') f(x,v'_1)-f (x,v)f(x,v_1)]\;,
\end{equation}
with
\begin{equation*} 
\label{scatt}
v'=v-n[n\cdot(v-v_1)] 
\end{equation*}
\begin{equation}\label{urto}
v_1'=v_1+n[n\cdot(v-v_1)]
\end{equation}
and $n$ a unit vector (impact vector) varying in
$S_+^2=\{n \in S^2\ |\ n\cdot(v-v_1)\geq 0\}$.  

Note that $v'$ and $v_1'$ are the outgoing velocities after a collision of
two elastic balls with incoming velocities $v$ and $v_1$ and centers $x$ and
$x+\varepsilon n$, with $\varepsilon$ the diameter of the spheres. The collision takes
place if $n\cdot(v-v_1) > 0$. Formulas \eqref{urto} are consequences of the conservation of energy and momenta.
Note that $\varepsilon$ does not enter \eqref{beq} as a parameter.
\begin{figure}
\begin{center}
\includegraphics[scale=0.16]{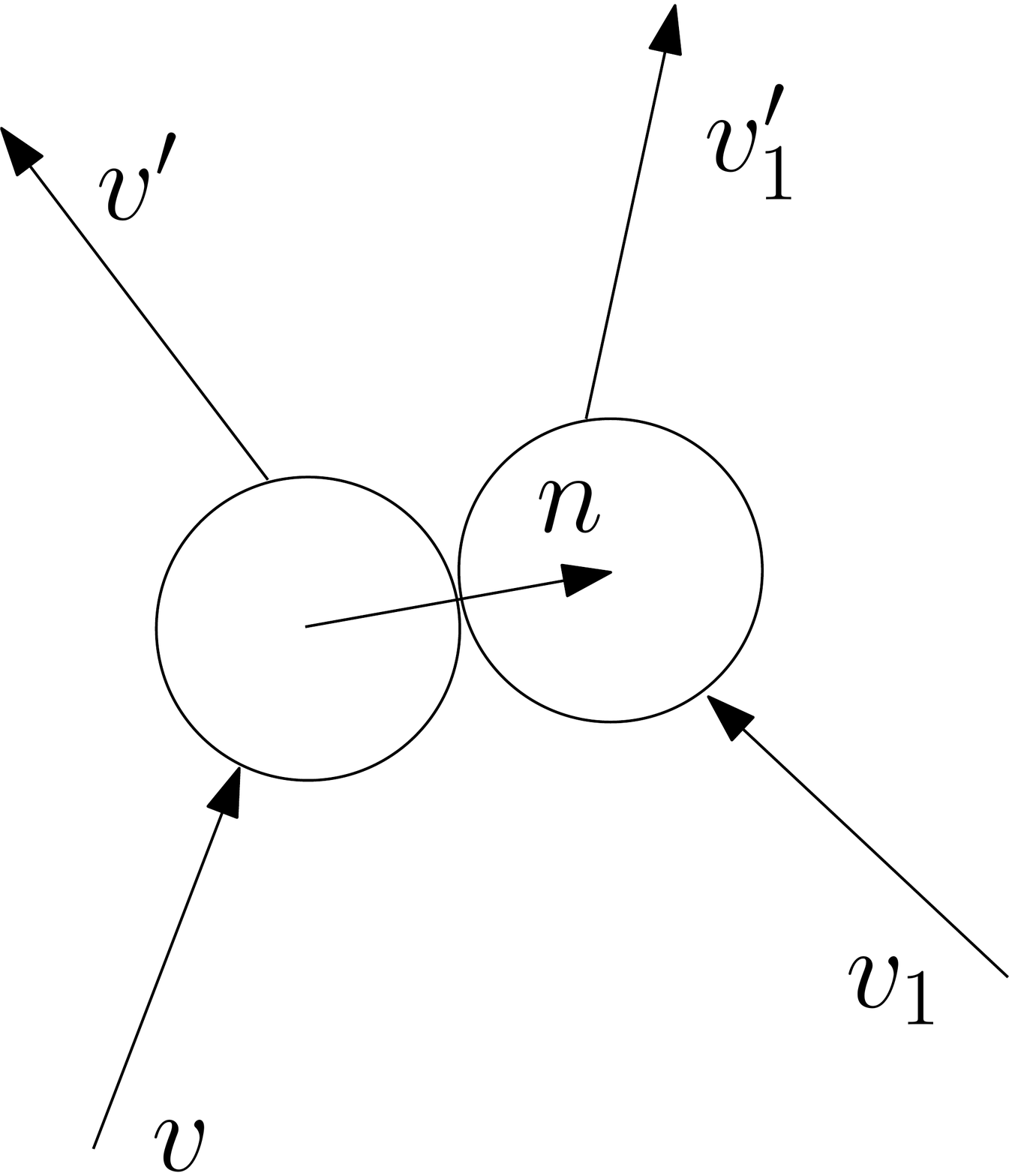}
\end{center}
\end{figure}

As a fundamental feature of \eqref{beq}, one has the formal conservation (in time) of the
five quantities
\begin{equation}\label{mass}
\int  dx  \int dv f(x,v,t) v^{\alpha}
\end{equation}
with $\alpha=0,1,2,$ expressing conservation of probability, momentum and
energy, respectively. From now on, we shall often abbreviate $\int=\int_{\mathbb{R}^3}$.

Moreover, Boltzmann introduced the (kinetic) entropy defined by 
\begin{equation}\label{entr}
H(f)=\int dx\int dv f\log f (x,v)
\end{equation}
and proved the famous $H$ theorem asserting the decrease of $H(f(t))$ along the
solutions of \eqref{beq}.

Finally, in the case of bounded domains or homogeneous solutions ($f=f(v,t)$
independent of $x$), the distribution defined for
some $\beta >0$, $\rho >0$ and $u\in \mathbb{R}^3$ by
\begin{equation}\label{max}
M(v)=\frac {\rho}{(2\pi / \beta)^{3/2}}e^{-\beta / 2 {|v-u|^2}},
\end{equation}
called Maxwellian distribution, is stationary for the evolution given by \eqref{beq}.
In addition, $M$ minimizes $H$ among all distributions with given total mass
$\rho$, mean velocity $u$ and mean energy. The parameter $\beta$ is interpreted
as the inverse temperature.

In conclusion, Boltzmann was able to introduce an evolution equation with the
remarkable properties of expressing mass, momentum and energy conservation and also
the tendency to thermal equilibrium. In this way, he tried to
conciliate Newton's laws with the second principle of thermodynamics. 

The $H$ Theorem is apparently in contrast with the  laws of mechanics,  which are time-reversible. This fact
caused  skepticism among the scientific community, and the work of Boltzmann was attacked repeatedly.
We refer the reader to the monograph by C.\;Cercignani \cite {C98}, which is
a beautiful compromise between historical account and scientific divulgation, to have a faithful idea
of the debate at the time.

To formally derive \eqref{beq},  let us consider a system of $N$ identical hard spheres of diameter $\varepsilon$  and
unitary mass, interacting by  means
of the collision law \eqref{urto}.  We denote by $\varepsilon$ the diameter of the particles which, for the moment, is fixed and not necessarily small.

The  phase space $\Gamma_N$ of the system is the subset of $\mathbb{R}^{3N}\times\mathbb{R}^{3N}$ fulfilling the hard-core condition, namely
$
  |x_i-x_j| \geq \varepsilon
$
for $i \neq j$.
The dynamical flow $Z_N\to \Phi^t (Z_N) $ is  defined as the free flow, i.e.,
$Z_N\to \Phi ^t (Z_N)= (x_1+v_1t ,v_1,\cdots ,x_N+v_Nt,v_N)$ up to the first impact time (when
$|x_i-x_j|=\varepsilon$); then an instantaneous collision takes place  according to the law \eqref{urto}, and the flow goes on up to the next collision instant. 

The well-posedness of the hard-sphere dynamics is not obvious, due to the occurrence of multiple collisions or to the a priori possibility that collision times accumulate at a finite limiting time. However, such pathologies  cannot occur outside a set of initial conditions  $Z_N$ of vanishing measure. Indeed following  \cite{Ale75} (see also \cite{CIP94}), the flow $Z_N\to \Phi ^t (Z_N) $ can be defined for all $t \in \mathbb{R}$ almost everywhere with respect to the Lebesgue measure, which is enough for what will follow (even the proof of this result is not relevant in the following, so that we omit further details).

Given a probability measure with density $W^N_0$ on $ \Gamma_N$,  thanks to the invariance of the Lebesgue measure under the above evolution, we define the time-evolved measure as the measure with density given by \eqref{meas}. 
Notice that this density is now restricted to $\Gamma_N$, however we can, equivalently and at any time, extend $W^N$ to zero outside $\Gamma_N$ and work with densities ``with holes'' in $ \mathbb{R}^{3N}\times\mathbb{R}^{3N}$.

We recall that we consider probability distributions $W^N$ which are initially (hence at any positive time)
symmetric in the exchange of the particles. 
The probability density of $j$ particles is then given by the $j$-particle marginal \eqref{marg}.

Note also that here $\Gamma_N, \Phi^t, W^N, f^N_j\cdots$ should exhibit a double dependence on $N$ and $\varepsilon$. We shall soon fix a precise $\varepsilon = \varepsilon(N)$ so that the notation becomes unambiguous. 

Cercignani \cite{Ce72} derived a hierarchy
of equations for the marginals  (in exactly the same spirit of the BBKGY hierarchy for smooth potentials), and  the first of such equations ($j=1$) is  
\begin{equation}\label {beq2}
(\partial_t+v\cdot \nabla_x)f^N_1=\hbox {Coll}\;,
\end{equation}
where $\hbox{Coll}$ denotes the variation of $f^N_1$ due to the collisions, which takes the form
\begin{equation}\label{bcoll2}
\hbox {Coll}=
(N-1)\,\varepsilon^2 \int dv_2 \int_{S^2} dn \,\, f^N_2 (x,v,x+n\varepsilon,v_2)\, (v_2-v)\cdot n\;.
\end{equation}
In the next section, we will comment on the justification of this equation. Here, let us accept it and
argue on its consequences.

%
%
%

Two given particles should be (almost) {\em uncorrelated} if the 
gas is {\em rarefied} enough. This leads to the propagation of chaos
\begin{equation}\label{caos}
f^N_2(x,v,x_2,v_2)\simeq f(x,v) f(x_2,v_2)\;,
\end{equation}
which might seem contradictory at first sight.
In fact, if two particles collide, correlations are created. Even assuming
\eqref{caos}  at some given time, if  particle $1$ collides with particle $2$,
such an equation cannot be satisfied at any time {\em after} the collision.

Before discussing the propagation of chaos further, we notice that, in practical situations,
for a rarefied gas, $N\varepsilon^3$
(total volume occupied by the
particles) is very small, while $N\varepsilon^2=O(1)$. This implies that the collision operator given by \eqref{bcoll2}
is $O(1)$.
Therefore, since we are dealing with a huge number of particles, we are 
tempted to perform the limit
$N\to \infty$ and $\varepsilon \to 0$ in such a way that
$\varepsilon^2=O(N^{-1})$. As a consequence,  
the probability that two tagged particles collide (which is of the order of the
surface of a ball, that is $O(\varepsilon^2)$), is negligible. Instead, the probability that 
a given particle collides with {\em any} of the remaining $N-1$
particles (which is $O(N\varepsilon^2)=O(1)$) is not negligible. On the other hand, condition \eqref {caos}
refers to two preselected particles (say $1$ and $2$) and it is not unreasonable to conceive that it holds in the limiting situation in which we work.  

Nevertheless, we cannot simply insert \eqref {caos} into \eqref {bcoll2}, as the integral operator
 refers to times both  before and after the collision. Let us assume \eqref {caos} only when the pair of velocities $(v,v_2)$
 are incoming ($ (v-v_2 ) \cdot n >0$).
If the two particles are initially uncorrelated, it is unlikely that they have collided before a given time $t$,
hence we assume  their  statistical independence.

This is a standard argument in textbooks of kinetic theory, but some extra care is needed.
 If particles $1$ and $2$  have not collided directly before a given time $t$,  this does not imply that they are uncorrelated. Indeed
there may exist a chain of collisions involving a group $ i_1,  i_2,  \cdots  $ of particles
 $$
 1 \to i_1 \to i_2 \to \cdots \to 2\;,
 $$
 correlating particles $1$ and $2$.
 As we shall see later, this is excluded (at least for a short time) by a more rigorous analysis. The two clusters of particles influencing the dynamics of particles $1$ and $2$ are disjoint with large probability.

Coming back to \eqref {bcoll2}, for the outgoing pair velocities $(v,v_2)$ (satisfying 
$(v_2-v)\cdot n>0$), we shall make use of the
continuity property
\begin{equation}\label{cont}
f^N_2 (x,v,x+n\varepsilon, v_2) =f^N_2 (x,v',x+n\varepsilon,v_2') \;,
\end{equation}
where the pair $(v',v_2')$ is  precollisional. On the two-particle distribution expressed in terms of 
precollisional variables, we apply now condition  \eqref {caos}, obtaining
\begin{equation}
\hbox{Coll}=(N-1)\varepsilon^2 \int dv_2 \int_{S_+^2} dn\,  (v-v_2)\cdot n
 [f (x,v') f(x-n\varepsilon,v'_2)-f (x,v)f(x+n\varepsilon,v_2)]
\end{equation}
after a change $n\to -n $   in the positive part of $\hbox{Coll}$ (remind the notation $S_+^2$
for the hemisphere  $\{ n \in S^2\ |\  (v-v_2)\cdot n> 0\} $). 

 Finally, in the limit $N\to \infty$ and $\varepsilon\to 0$ with $N\varepsilon^2=\lambda^{-1}>0$, we find:
\begin{equation}
\label{beq1}
(\partial_t+v\cdot \nabla_x)f=
\lambda^{-1} \int dv_2 \int_{S_+} dn \, (v-v_2)\cdot n \, [f (x,v') 
f(x,v'_2)-f (x,v)f(x,v_2)].
\end{equation}
The parameter $\lambda$ represents, roughly, 
the typical length a particle can cover without undergoing any collision ({\em mean free path}). (In \eqref {bcoll},
 we just chose $\lambda=1$.)

It may be worth remarking that, after having taken the limit
 $N \to \infty$ and $\varepsilon \to 0$, there is no way to distinguish between incoming and outgoing pair velocities.
 This is because no trace of  the parameter $\varepsilon$ is left in \eqref{beq1} and $n$ plays the role of a random variable. 
 However, keeping in mind the way the Boltzmann equation was derived, one shall conventionally maintain the name
 {\em incoming} for velocities satisfying the condition $(v-v_2 )\cdot n > 0$ (and consequently the pair $(v',v'_2)$
 would be outgoing in \eqref{beq1}).

Equation \eqref {beq1} (or  equivalently \eqref {beq}-\eqref {bcoll}) is the Boltzmann
equation for hard spheres. Such an equation has a statistical nature, and it is 
not equivalent to the hamiltonian dynamics from which it has been derived. Indeed
the $H$ theorem shows that it is not reversible in time in contrast with the laws of
mechanics.

By the analysis on the order of magnitude of the quantities in the game, we deduced that the Boltzmann equation
works in special situations only. The condition $N \varepsilon^2 =O(1)$ means that we consider a rarefied gas,
with almost vanishing volume density.
After Boltzmann established the equation, Harold Grad \cite{Gr49,Gr58} postulated its  
validity in the limit $N \to \infty$ and $ \varepsilon \to 0$ with $N \varepsilon^2 \to O(1)$ as discussed above
(this is often called, indeed, the Boltzmann-Grad limit).

There is no contradiction in the irreversibility or in the trend to equilibrium obtained after the limit,
when they are strictly speaking false for mechanical systems.
However, the arguments above are delicate and require a rigorous,
deeper analysis. If the Boltzmann equation is not a purely
phenomenological model derived by assumptions ad hoc and justified by
its practical relevance, but rather a {\em consequence} of a mechanical model, we
must derive it rigorously. In particular, the propagation of chaos should not be
a hypothesis but the statement of a theorem.

After the formulation of the mathematical validity problem by Grad, Cercignani \cite{Ce72}
obtained the evolution equation (hierarchy) for the marginals of a hard-sphere system,
and this was the starting point to rigorously derive the Boltzmann equation,
as accomplished by Lanford in his famous paper \cite {La75}, even though only for a
short time interval.

Lanford's theorem is probably the most relevant result regarding the mathematical
foundations of kinetic theory. In fact, it dispelled the many previous doubts
on the validity of the Boltzmann equation (although some authors refuse a priori
the problem of deriving the equation starting from mechanical systems \cite{TM}).

Unfortunately, the short-time limitation is a serious one. Only for special systems, as
is the case of a very rarefied gas expanding in a vacuum, can we obtain a global
validity \cite{IP86,IP89}. The possibility of deriving the
Boltzmann equation globally in time, at least in cases when we have a global
existence of good solutions, is still an open, challenging problem.

We conclude this section with a few historical remarks.
Before Boltzmann, Maxwell proposed a kinetic equation that is just the Boltzmann equation integrated against test functions  \cite {Ma67,Ma95}.
He considered also more general potentials, in particular, inverse-power-law potentials,
motivated essentially by the special properties of their cross-section.
After Lanford's result, the case of smooth short-range potentials has been studied
by other authors \cite{Ki75,GSRT12,PSS13}. It is a nontrivial extension, in particular when the
interacting potential is not ``close enough'' to  a hard-sphere potential.
The validity (or nonvalidity) of the Boltzmann equation in the case of genuine long-range
interactions is open, in absence of techniques suited to deal with collisional and mean-field terms 
simultaneously.

\subsection{Hard-sphere hierarchies}
\label{sec:2.1}

In this and in the following section we give more details on the derivation of \eqref{beq} from $N$ hard spheres of diameter $\varepsilon$, discussed above heuristically. We remind the reader that we are interested in the behaviour of the system in the limit $N\to \infty,  \varepsilon \to 0$ fixing $ \varepsilon ^2N =1$ ($1$ chosen for simplicity), according to the Boltzmann-Grad limit.  Namely we have a single scaling parameter $ \varepsilon$ (or $N$), and we study the asymptotics $ \varepsilon \to0$ ($N \to \infty$).

We start with the justification of \eqref{meas}. Let  $A$ be a measurable set in $\mathbb{R}^{3N}\times \mathbb{R}^{3N}$. Then the probability  of finding the system in $A$ at time $t>0$ is  given by
$$
\mathbf{P}_t (A) =\mathbf{P}_0 (\Phi^{-t} (A) )
$$
where
$$
\Phi^{-t} (A)=\{ Z_N \ |\  \Phi^t (Z_N) \in A \}
$$
(dropping the dependence on $N = \varepsilon^{-2}$).
If $\chi_A$ is the characteristic function of $A$, we have that
$$
\int W^N(Z_N,t) \chi_A(Z_N) =\int W^N_0(Z_N) \chi_{\Phi^{-t} (A)} (Z_N)= \int W^N_0(Z_N) \chi_{A} (\Phi^{t}(Z_N))\;,
$$
which implies that
\begin{equation}
\int W^N(Z_N,t)\, u(Z_N) =\int W^N_0(Z_N) \, u (\Phi^{t}(Z_N))
\label{eq:leqd}
\end{equation}
for any bounded Borel  function $u$.  Here the integral is extended over all the phase space $\Gamma_N$. By using the Liouville theorem on the transformation $Z_N \to \Phi^{t}(Z_N)$,  it follows that
$$
 W^N(  \Phi^{t}(Z_N), t) =W^N_0(Z_N)\;,
$$
or  \eqref{meas} by the invertibility of the same transformation. 

This probability distribution is not expected to converge. Thus, we focus immediately on the collection of marginal distributions $(f^N_j)_{j \geq 1}$, given by \eqref{marg}, for which the evolution equation has the form
\begin{equation}\label{hie2}
(\partial_t+{\cal L}_j^\varepsilon )f _j^N =(N-j) \,\varepsilon^2 \, C^ \varepsilon_{j+1} f_{j+1}^N\;,\quad j=1,\cdots,N-1\;.
 \end{equation}
Here  ${\cal L}_j^\varepsilon $ is the generator of the dynamics of $j$ hard spheres of diameter
$\varepsilon$ (Liouville operator of a $j-$particle system), while
 \begin{equation}\label{hcollsum}
C^\varepsilon_{j+1}=\sum_{k=1}^j C^ \varepsilon_{k,j+1}\;,
 \end{equation}
  \begin{equation}
\label{hcollk}
C^ \varepsilon_{k,j+1}f^N_{j+1}(Z_j)= \int dv_{j+1} \int_{S^2} dn \, (v_{j+1}-v_k)\cdot n \,\,
  f^N_{j+1}(Z_j, x_{k}+\varepsilon n,v_{j+1})
 \end{equation}
%
 is the $j$-particle collision operator (generalizing \eqref{bcoll2} to higher orders).
 For $j=N$, we are left with the Liouville equation in a differential form, namely
 $f^N_N=W^N$ and 
 \begin{equation}
 \label{eq:l2}
 (\partial_t +{\cal L}^\varepsilon _N ) W^N=0\;.
 \end{equation}

To derive Eq.\,\eqref{hie2} formally, we would like to give some description of $ {\cal L}^ \varepsilon_j$ as differential operator. This poses a difficulty, in fact ${\cal L}^\varepsilon_j = \sum_{i=1}^j v_i \cdot \nabla_{x_i} $ on functions vanishing on $\partial\Gamma_j$ and the interacting dynamics is completely coded on the boundary. In \cite {Ce72,CIP94}, boundary conditions are
imposed using \eqref{cont}, and its higher order versions, and Eq.\,\eqref{hie2} is derived integrating by parts over $\Gamma_N$.
However if one is not afraid of working with delta functions, it is more convenient to use the following compact description:
\begin{equation}
\label{gen}
{\cal L}^\varepsilon_j = \sum_{i=1}^j v_i \cdot \nabla_{x_i}- {\cal T}^{\varepsilon}_j
 \end{equation}
where
\begin{equation}
\label{gen1}
{\cal T}^{\varepsilon}_j=  \sum_{\substack{i<k \\ i,k=1, \cdots, j }} {\cal T}_j^{\varepsilon;\, i,k}
 \end{equation}
and
\begin{equation}
\label{gen2}
{\cal T}^{\varepsilon; \,i,k}_j f^N_j  (Z_j)= \varepsilon^2 \int_{ S^2}  dn  \left(U_{i,k}\cdot n \right)_+
\big [ \delta (R_{i,k} -  \varepsilon n) b_n^{i,k}- \delta (R_{i,k} +  \varepsilon n) \big ] f^N_j (Z_j)\,,
 \end{equation}
with $U_{i,k}=v_i-v_k,\; R_{i,k}=x_i-x_k $, $(\cdot)_+$ is the positive part, and
$$
b_n^{i,k} f^N_j (Z_j)=  f^N_j(x_1, v_1 \cdots x_i,v'_i \cdots x_k,v'_k \cdots x_j,v_j)\;.
$$
The last operator transforms the incoming pair $(v_i,v_k)$ into the outgoing $(v'_i,v'_k)$ after a scattering with impact parameter $n$. 
Note finally that the operator \eqref{hcollk} can be as well expressed in terms of ${\cal T}^{\varepsilon}_j$:
 \begin{equation} \label{eq:C-T}
\varepsilon^2 C^ \varepsilon_{k,j+1}f^N_{j+1}= \int dz_{j+1} {\cal T}^{\varepsilon; \,k, j+1}_{j+1}
  f^N_{j+1}\;.
 \end{equation}
 We should remind here that  the marginals are supported on the space of non-overlapping hard spheres ($|x_i-x_k| \geq \varepsilon$ for $i \neq k$). Therefore, when we think of \eqref{hie2} (and \eqref{eq:l2}) as equations over the whole space $\mathbb{R}^{3j}\times \mathbb{R}^{3j}$, we should always complement them with the condition $|x_i-x_k| < \varepsilon \Rightarrow f^N_j = 0$.

Let us now check the expression given for the Liouville equation, based on ${\cal T}^{\varepsilon}_N$.
Consider a point particle hitting a sphere of diameter $ \varepsilon$ of infinite mass, centred at the origin. Let $g(X,V,t)$ be the probability distribution of the point particle, with initial datum $g(t=0) = g_0$.
Let $V$ and $V'$ denote the incoming and outgoing velocity, respectively. It is $ V'=V-2(V\cdot n) n$, where $n\in S^2$ is the impact vector.
\begin{figure}
\begin{center}
\includegraphics[scale=0.5]{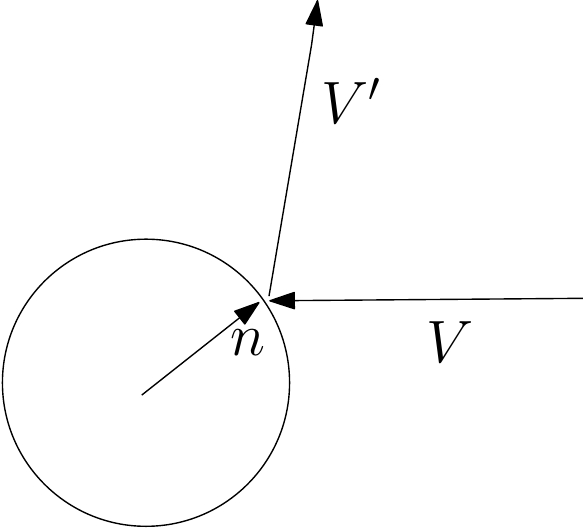}
\end{center}
\end{figure}

\noindent
We denote by $Z=(X,V) \to Z(t) = (X(t),V(t))$ the dynamical flow. For any test function $\varphi=\varphi(X,V)$ we have that
$$
\frac d {dt} \varphi(Z(t) )= V\cdot \nabla_X \varphi( Z(t))+ \delta (t-\tau) [ \varphi(X+V \tau,V')-  \varphi(X+V \tau,V) ]
$$
where $\tau$ is the hitting time. 
The term $[ \cdots]$ describes the jump in velocity. 
Proceeding as in \eqref{eq:leqd} we deduce that
\begin{equation}
\label{d}
\frac d {dt} \iint g(Z,t)\, \varphi(Z)
= -  \iint V\cdot \nabla_X  g(Z,t) \,\varphi(Z )+
\int_{A_{coll}} g_0(Z) \,\delta (t-\tau)\, [ \cdots ].
 \end{equation}
Here  $A_{coll}$ is the  set of configurations $Z$ delivering a collision in the 
future. Introducing the change of variables
$$
(X,V) \in A_{coll} \to (n \varepsilon, \tau, V)\;,
$$
which has jacobian determinant of modulus $\varepsilon^2 |V \cdot n  |$, the last term in Eq.\,\eqref{d} becomes
\begin{eqnarray}
&& \varepsilon^2 \int_0^\infty d\tau \int dV \int_{ V \cdot n < 0} dn \, |V \cdot n |\, g_0(Z(n \varepsilon, \tau, V))\, \delta(t-\tau) 	\, [ \varphi(n \varepsilon, V')-  \varphi(n \varepsilon ,V) ] \nonumber \\
&& = \varepsilon^2  \int dV \int_{ V \cdot n < 0} dn\iint d\tilde Z  |V \cdot n | g_0(\tilde Z)\delta(\tilde Z-Z(n \varepsilon, t, V))\,  [ \varphi(n \varepsilon, V')-  \varphi(n \varepsilon ,V) ] \nonumber \\
&& = \varepsilon^2  \int dV \int_{ V \cdot n < 0} dn\iint d\tilde Z\,  |V \cdot n | \,g(\tilde Z,t)\,\delta(\tilde Z-(n \varepsilon,V))\,  [ \varphi(n \varepsilon, V')-  \varphi(n \varepsilon ,V) ] \nonumber \\
&& =  \varepsilon^2 \, \int dV\Big[ \int_{V \cdot n > 0} dn  \,|V \cdot n|  \,g (n \varepsilon, V',t)- \int_{V \cdot n < 0} dn  \,|V \cdot n|  \,g (n \varepsilon, V,t)\Big] \,\varphi(n \varepsilon, V) \;.\nonumber
\end{eqnarray}
Note that in the last step we changed again variables, $V \to V'$, in the positive term. To identify the time derivative in strong form, we now write $g(n\varepsilon,\cdot)\varphi(n \varepsilon, \cdot) = \int dX \delta(X-n\varepsilon)
g(X,\cdot)\varphi(X, \cdot)$, exchange the integrals. and make a last change of variables $n \to -n$ in the negative term. We conclude that
$$
\frac d {dt} g(Z,t) =-  V\cdot \nabla_X  g(Z,t) +  \varepsilon^2 \int_{ V \cdot n > 0} dn\,  |V \cdot n | \,
[\delta( x- \varepsilon n)\,b_n -\delta(x+ \varepsilon n)]\, g(Z,t)\;,
$$
where $b_n$ flips $V$ into $V'$. The general form of ${\cal T}^{\varepsilon}_N$ follows easily from this computation.

In order to arrive to \eqref{hie2}, it is enough to proceed as in the standard derivation of the BBGKY hierarchy for smooth potentials. We split the sums in ${\cal T}^{\varepsilon}_N$ given by Eq.\,\eqref{gen1}, as
$$
 \sum_{\substack{i<k \\ i,k=1, \cdots, N }} = \sum_{i<k \leq j } + \sum_{i \leq j, k>j } + \sum_{j<i<k }\;,
$$
and integrate in $dz_{j+1} \cdots dz_N$. The first sum produces ${\cal L}^\varepsilon_j$ immediately. The second sum gives the collision operator $\varepsilon^2 C^\varepsilon_{j+1}$, multiplied by a symmetry factor $(N-j)$. The last sum vanishes by exact compensation of gain and loss in \eqref{gen2}.

Eq.\,\eqref{hie2} is the starting point for Lanford's validity theorem, as we shall see in the following section.

\subsection{Lanford's Theorem}

The iteration of the Duhamel formula for Eq.\,\eqref{hie2}
leads to express $f^N_j(t)$ as a sum:
\begin{eqnarray}
\label{series}
f^N_j(t)=&& \sum_{n=0}^{N-j} \alpha_n (N,j) \int_0^t dt_1 \int _0^{t_1} dt_2 \dots \int_0^{t_{n-1}} dt_n  \\
&&S^ \varepsilon (t-t_1) \,C^ \varepsilon_{j+1}\, S^ \varepsilon (t_1-t_2) \cdots S^\varepsilon (t_n)\, f^N_{j+n} (0)\,,\nonumber
 \end{eqnarray} 
where $S^ \varepsilon (t) F(Z_j)=F(\Phi^{-t} Z_j)$ is the $j$-particle interacting flow and 
$$
\alpha_n(N,j)= (N-j) \cdots (N-j-n+1)  \,\varepsilon^{2n}.
$$
The main ingredient for the theorem of Lanford stated below, is just this explicit representation for the solution of the $N$-particle hierarchy. Actually, this identity can be rigorously proved directly, without making use of \eqref{hie2} \cite{Si13,PS15}. 

On the other hand, a similar formula can be established for the tensor product of solutions to the Boltzmann equation $f_j(t):=f(t)^{\otimes j}$. Namely we have that
\begin{eqnarray}
\label{seriesB}
f_j(t)=&& \sum_{n=0}^{\infty}  \int_0^t dt_1 \int _0^{t_1} dt_2 \dots \int_0^{t_{n-1}} dt_n  \\
&&S (t-t_1)\, C_{j+1} \,S (t_1-t_2) \cdots S (t_n) \,f_{j+n} (0) \nonumber.
 \end{eqnarray} 
Here $S(t) F(X_j,V_j)=F(X_j-V_jt,V_j)$ is the free-flow operator, and $C_{j+1}=\sum_{k=1}^j C_{k,j+1}$, where $C_{k,j+1}$ is the formal limit for $ \varepsilon \to 0$ of \eqref{eq:C-T}.
%

Since $\alpha_n(N,j) \to 1$ and $S^ \varepsilon(t) \to S(t)$ almost everywhere in the limit, each term in the right-hand side of  \eqref{seriesB} is the limit 
of the corresponding term in  \eqref {series}, provided that we require  a good behaviour of the initial datum $f^N_{j+n} (0)$.

We cannot simply require that $f^N_{j+n} (0)=f_j(0)$, because the hard-core condition induces correlations at time zero.
Let  $f_0$ be a one-particle probability distribution, and the initial datum for the Boltzmann equation. We make the following assumptions.

\smallskip
{\bf Hypothesis 1. }
$f_0 \in C(\mathbb{R}^6 \to \mathbb{R}^+)$, $\int f_0 =1$. Moreover
$$
f_0(x,v) \leq h(x) e^{-\beta v^2}
$$
where $h \in L^1 (\mathbb{R}^3) \cap L^\infty (\mathbb{R}^3)$, $\| h \|_{L^\infty} =z_0$ and $\beta >0$.

\smallskip
{\bf Hypothesis 2. }
Let $\Gamma_j^{\neq}$ be the subset of $\mathbb{R}^{3j}\times \mathbb{R}^{3j}$ fulfilling the condition
$$
x_i \neq x_k \quad \text {for }\quad i \neq k, \, 1 \leq i,k \leq j.
$$
Then, the marginals of the hard-sphere system satisfy
 \begin{equation}
\label{conv0}
\lim_{ \varepsilon \to 0} f^N_{ j}(t=0) = f_0^{\otimes j}\;,
  \end{equation}
uniformly on compact subsets of $\Gamma_j^{\neq}$ and the bound
$$
 f^N_{ j} (x_1, v_1, \cdots x_j, v_j, t=0) \leq
\prod_{i=1}^j h(x_i) e^{-\beta v_i^2}.
$$
\smallskip
{\bf Theorem \cite{La75}. }
 {\it Under the hypotheses 1 and 2, there exists $t_0 >0$ (depending only on $z_0,\beta$) such that, for $t<t_0$
we have, for all $j\geq 1$,
 \begin{equation}
\label{convt}
\lim_{ \varepsilon \to 0} f^N_j (t)=f (t)^{\otimes j}
 \end{equation}
where $f(t)$ is the unique solution to the Boltzmann equation. The convergence holds almost everywhere.
}

\smallskip

Following Lanford, the proof can be organized in two steps.

We first give an a priori bound on  the series expansions   \eqref{series} (uniform in $\varepsilon$)
  and  \eqref{seriesB}, using that the time $t$ is small enough.
  To give a rough idea of this step, let us cutoff large velocities. In particular, we ignore the factors $|v_{j+1}-v_k|$ in \eqref{hcollk}. Then, the string of operators can be estimated brutally by
    $$
 |S^ \varepsilon (t-t_1) \,C^ \varepsilon_{j+1}\, S^ \varepsilon (t_1-t_2) \cdots S^\varepsilon (t_n)\, f^N_{j+n} (0)|
 \leq  C^{j+n} j(j+1) \cdots (j+n-1)
  $$
 for some $C > 0$, where the factorial growth comes from the sum in \eqref{hcollsum}.
On the other hand, the ordered time-integration yields $t^n / n!$, so that the series expansion is bounded by a geometric series
$
\sum_n C_1^j (C_2 t)^n\;,
$
for positive $C_1, C_2$. 

In the second step, one shows the term by term convergence of  \eqref{series} to  \eqref{seriesB}. Here the short time restriction does not enter anymore.

For more details on Lanford's proof, we refer to \cite{La75,Sp91,CIP94}.

We conclude with some remarks.
\begin{enumerate}
\item The time $t_0$ is explicitly computable. It turns out to be a fraction of
the mean free time between collisions. This time limitation is purely technical.
\item  Lanford's original proof was qualitative: it does not make explicit the rate of convergence. This can be obtained
with some extra care, along the same arguments \cite {GSRT12, PSS13}.
\item Initial conditions fulfilling Hypotheses 1 and 2 can be easily constructed. The most natural
initial state is maximally factorized, meaning that the only source of correlation is due to the hard-core exclusion.
In this case, the $N$-particle measure is
 \begin{equation}
W^N_0(Z_N):=\frac{1}{\mathcal{Z}_N}f^{\otimes N}_0(Z_N) \prod_{1\leq i < k \leq N}{ \bf 1}_{\{|x_i-x_k|> \varepsilon\}}(Z_N)\;,
\label{eq:exampleWN0}\nonumber
 \end{equation}
where
 \begin{equation}
\mathcal{Z}_N:=\int_{\mathbb{R}^{3N}\times\mathbb{R}^{3N}} dZ_N f^{\otimes N}_0(Z_N)\prod_{1\leq i < k \leq N} {\bf 1}_{ \{|x_i-x_k|> \varepsilon\} }(Z_N)\nonumber
 \end{equation}
is a normalization factor, and $f_0$ satisfies Hypothesis 1. For this state, the verification of \eqref{conv0} is a simple exercise.
\label{sec:3}
\end{enumerate}

\section{Weak-coupling limit and Landau equation}
\label{sec:3}

The Boltzmann equation is suited to the description of rarefied gases, and one can ask
whether a useful kinetic analysis can be applied also to the case of a {\em dense} gas. 
To introduce the problem, let us revisit the Boltzmann-Grad limit in an alternative
way. Let $ \varepsilon$ be a small scale parameter  denoting the ratio between the microscopic and
the macroscopic scale, for instance the inverse number of atomic diameters necessary to cover $1$ meter, or
the inverse number of atomic characteristic times necessary to cover $1$ second.
Then, scale space and time by $ \varepsilon$ in the equations of motion (in our case, the hard-sphere hierarchy).
 We need to specify the number of particles $N$. In a box of side $1$, 
there should be $N\approx  \varepsilon^{-3}$ particles if one assumes that the intermolecular distance is of the same order of the molecular diameter. The number of collisions of a given particle per macroscopic unit time would be $ \varepsilon^{-1}$.  As we have seen, in a low-density regime, $N$ scales differently, namely $N\approx  \varepsilon^{-2}$, 
the number of collisions per unit time is finite and the one-particle distribution function satisfies the Boltzmann equation.  

A variety of possible scalings describes different physical situations (see the next section).
For instance, the gas may be dense, $N=O( \varepsilon^{-3})$ and the particles are {\em weakly
interacting} via a smooth two-body potential $\phi$.  
To express the weakness of the interaction, we assume
that $\phi$ is rescaled by $\sqrt { \varepsilon}$. 
Since $\phi$ varies on a scale $ \varepsilon$ (in macroscopic unities), the force
will be $O(\frac 1{\sqrt { \varepsilon}})$ and act on a time interval $O( \varepsilon)$.
The variation of momentum due to the single scattering is $O(\sqrt { \varepsilon})$, and the
number of particles met by a typical particle is $O(\frac 1{ \varepsilon})$. Hence, the
total momentum variation for unit time is $O(\frac 1{\sqrt { \varepsilon}})$.
However, in the case of a homogeneous gas and symmetric forces, this variation should be
zero in the average. The computation of the variance leads to a result $\frac
1{ \varepsilon} O(\sqrt { \varepsilon})^2=O(1)$. Therefore, based on a {\em central-limit} type
of argument, we expect that in the kinetic limit a diffusion equation in the velocity variable holds.

At the level of kinetic equations,
consider a collision operator of Boltzmann type, for a spherically symmetric, smooth potential $\phi=\phi(x)$.
We assume for simplicity the potential to be short-range, namely $\phi(x)=0$ if $|x|>1$.
The collision operator $Q_B$ is given by \eqref{bcoll}, with \eqref{urto} replaced by
\begin{equation*} 
\label{scatt}
v'=v-\omega[\omega\cdot(v-v_1)] 
\end{equation*}
\begin{equation}\label{urto'}
v_1'=v_1+\omega[\omega\cdot(v-v_1)]
\end{equation}
where $\omega$ is the unit vector in the direction of the transferred momentum, while $n$ is the impact parameter\footnote{Note that this is not the conventional form for the Boltzmann equation and usually the factor
$(v-v_1) \cdot n $ is rewritten in terms of $\omega$, which amounts to introduce the differential cross-section.}. 
The potential $\phi$ enters in the determination of $\omega$.
\begin{figure}
\begin{center}
\includegraphics[scale=0.7]{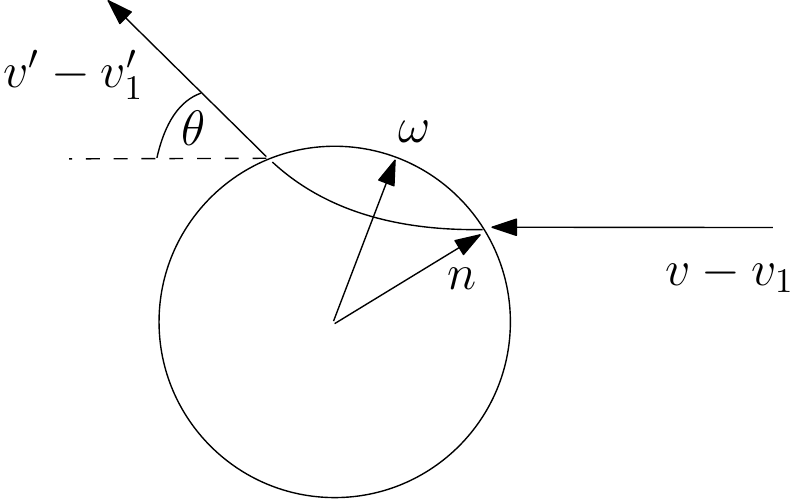}
\end{center}
\end{figure}

According to the weak-coupling-limit prescription discussed above, we rescale the potential as $\phi_ \varepsilon= \sqrt{ \varepsilon } \phi (\frac x { \varepsilon})$, and simultaneously increase the density. The new collision operator reads 
\begin{equation}
\label{Bsmooth}
Q_B^\varepsilon(f,f)(x,v)=  \frac 1{ \varepsilon} \int d v_1 \int_{S_+^2}  \,\, d n \, (v-v_1)\cdot n 
\,\, \{f(x,v+p)f(x,v_1-p)-  f(x,v)f(x,v_1)\}
 \end{equation}
where $p=-\omega\cdot (v-v_1)\, \omega$ is the transferred momentum, which is  typically $O( \sqrt{ \varepsilon})$.

It follows that, for any smooth test function $u=u(v)$, setting $U=v-v_1$ (omitting the spatial dependence),
\begin{eqnarray}
\label{exp}
&& \int dv\, u(v)\,Q_B^\varepsilon (f,f)(v)\nonumber\\&& = \frac 1{2 \varepsilon} \iint  dv dv_1 \int_{S_+^2}  \,\, d n \, U \cdot n 
\, \Big\{ u(v+p)+u(v_1-p) - u(v )-u(v_1) \Big\}\,f(v) f(v_1)\nonumber \\
 && \approx 
 \frac 1{2 \varepsilon} \iint  dv dv_1 \int_{S_+^2}  \,\, d n \,U \cdot n \,\,\,
\Big\{ p\cdot (\nabla_v u(v)-\nabla_{v_1} u(v_1))+  \\
&&\ \ \ \  \sum_{\alpha,\alpha'} \left( \frac 12 \partial^2_{\alpha,\alpha'}u(v) p_{\alpha} p_{\alpha'} +\frac 12 \partial^2_{\alpha,\alpha'}u(v_1) p_{\alpha} p_{\alpha'}\right)  \Big\}         
f(v) f(v_1)\,, \nonumber 
 \end{eqnarray}
where we Taylor-expanded up to second order in $p$ in order to compensate the divergence $ \frac 1{ \varepsilon}$, and the $\alpha$'s run over the three vector components.

We first analyze the second order. 
Let $x(s)$ be the trajectory of one particle scattering in the central potential $\phi_ \varepsilon$
with incoming velocity $U$ and initial time fixed by $x(0)= \varepsilon n$.
To evaluate
$$
T_{\alpha,\alpha'}(U):= \frac 1{2  \varepsilon} \int_{S_+^2}  \,\, d n\,  U \cdot n \,\, p_{\alpha}p_{\alpha'}\;,
$$
we write
$$
p_{\alpha} =-\int_{-\infty}^{+\infty} ds \,\frac 1 {\sqrt { \varepsilon} } \,\nabla_{x_\alpha}\phi \left(\frac {x(s)}{ \varepsilon}\right) =- \left(\frac 1{ \sqrt{2\pi}}\right)^3 \frac 1 {\sqrt { \varepsilon}} \int ds \int_{\mathbb{R}^3} dk
\, \mbox{i}\,k_\alpha e^{ \mbox{i}\, k \cdot \frac {x(s)} { \varepsilon }} \hat \phi(k)\;.
$$
Then
\begin{eqnarray}
&&T_{\alpha,\alpha'}=-  \left(\frac 1{2\pi}\right)^3 \frac 1{2  \varepsilon^2 } \int_{S_+^2}  \,\, d n\,  U\cdot n \nonumber\\
&&\ \ \ \  \int ds_1 \int ds_2 \int dk_1 \int dk_2 \,\, ( k_1)_\alpha (k_2)_{\alpha'} \,\, e^{ \mbox{i} k_1 \cdot \frac {x(s_1)} { \varepsilon }} \,\, e^{ \mbox{i} k_2 \cdot \frac {x(s_2)} { \varepsilon }} \, \hat \phi(k_1)  \hat \phi(k_2). \nonumber
 \end{eqnarray}
But $ \frac {x(s)}{ \varepsilon} \approx n+\frac {Us} { \varepsilon}$. Therefore, setting $y(s)=n+Us$ (after rescaling times) we have that
\begin{eqnarray}
&&T_{\alpha,\alpha'}=-  \left(\frac 1{2\pi}\right)^3 \frac 1{2  } \int_{S_+^2}  \,\, d n\,  U\cdot n \nonumber\\
&&\ \ \ \  \int ds_1 \int ds_2 \int dk_1 \int dk_2 \,\, ( k_1)_\alpha (k_2)_{\alpha'} \,\, e^{ \mbox{i} k_1 \cdot y(s_1)} \,\, e^{ \mbox{i} k_2 \cdot y(s_2)} \, \hat \phi(k_1)  \hat \phi(k_2). \nonumber
 \end{eqnarray}
Next we write
$$
e^{ \mbox{i} k_1 \cdot y(s_1)} \,\, e^{ \mbox{i} k_2 \cdot y (s_2)}= e^{ \mbox{i} (k_1+k_2) \cdot y(s_1)} \,\, e^{ \mbox{i} k_2 \cdot U(s_2-s_1)}
$$
and change variables in the following way. Setting
$
\tau =s_2-s_1,
$
parametrize the points $\xi$ of the cylinder with axis $-U/|U| $ and basis the unit circle through the origin, by
$(n,s_1) \to n+Us_1 $. Then
$
d\xi= d n \, ds_1\,(U \cdot n)_+ 
$
and
\begin{equation}
\label{T}
T_{\alpha,\alpha'} \approx - \left(\frac 1{2\pi}\right)^3 \frac 1{2 } \int d \xi \,d\tau\, dk_1 \,dk_2 \,( k_1)_\alpha (k_2)_{\alpha'} e^{ \mbox{i} (k_1+k_2) \cdot \xi} e^{ \mbox{i} k_2 \cdot U \tau}
\hat \phi(k_1)  \hat \phi(k_2)\;,
 \end{equation}
hence we arrive to
\begin{equation}
\label{Tfinal}
T_{\alpha,\alpha'} \approx  -\frac {(2\pi )}{2 } \int dk \,\hat \phi^2(k) \,\delta ( k  \cdot U)\, k_\alpha k_{\alpha'} =: a_{\alpha,\alpha'}(U)\;.
 \end{equation}
This matrix can be handled conveniently  by means of polar
coordinates $k=\hat k \rho$, $\hat k=\frac {k} {|k|}$:
\begin{equation}
a_{\alpha,\alpha'}(U)= -\frac {(2\pi )^4}{2 } \frac 1{|U |} \int d\rho \,\hat \phi^2 (\rho) \,\rho^3 \int d \hat k \;  \; \delta (\hat U \cdot
\hat k)\;,
 \end{equation}
where $\hat U$ is the versor of $U$ . Here we are using that, due to the spherical symmetry, $\hat \phi$  depends on $k$ through $|k|$ only. 
Setting
\begin{equation}
B=\pi \int_0^{\infty}d\rho\, \hat \phi^2 (\rho) \,\rho^3 
 \end{equation}
and computing $\int d \hat k\, \delta (\hat U \cdot
\hat k)$, we conclude that
\begin{equation}
\label{a}
a_{\alpha,\alpha'}(U)=\frac B{|U|} \left(\delta_{\alpha,\alpha'}-\hat U_\alpha \hat U_{\alpha'}\right)\;.
 \end{equation}
$B$ is the kinetic constant coding all the information on the microscopic potential. 

We turn now to the evaluation of the first order terms in \eqref{exp}, i.e.
$$
T(U):= \frac 1{2  \varepsilon} \int_{S_+^2}  \,\, d n  \,U \cdot n \,\, p_U, \qquad  T_{\perp}(U):= \frac 1{2  \varepsilon} \int_{S_+^2}  \,\, d n \, U \cdot n \,\, p^{\perp}_U
$$ 
where $p=(p_U,  p_U^{\perp} )$ and $p_U>0$ is the projection of $p$ over $-U/|U|$.
Note that $T_{\perp}$ is vanishing by symmetry.
On the other hand, $p_U=(\omega\cdot U)^2 / |U| = p^2/ |U|$
so that
$$
T(U) =   \frac 1{|U|} \sum_\alpha T_{\alpha,\alpha}(U) \approx  \frac 1{|U|}\sum_\alpha a_{\alpha,\alpha} (U)
= \frac {2B}{|U|^2}\;.
$$ 

In conclusion,
\begin{equation}
\label{weak}
 \int dv\, u(v)\,Q_B^\varepsilon (f,f)(v) \approx \iint dvdv_1 \,L u(v,v_1)\,f(v) f(v_1) 
 \end{equation}
where
$$
Lu(v,v_1):=-2 \frac B {|v-v_1|^3}(v-v_1) \cdot (\nabla_v u(v)-\nabla_{v_1} u(v_1))+Tr(a\otimes D^2u)(v,v_1)
$$
where   $Tr(a\otimes D^2u)(v,v_1) =\sum_{\alpha,\alpha'}a_{\alpha,\alpha'} (v-v_1)\partial^2_{ v_\alpha,v_{\alpha'}} u(v)$ and $a = (a_{\alpha,\alpha'})_{\alpha,\alpha'}$ is given by  \eqref{a}.

This leads to introduce the Landau operator, defined by
\begin{equation} \label{eq:LO}
Q_L(f,f)(x,v):=\int dv_1 \nabla_v \; a(v-v_1) \,(\nabla_v-\nabla_{v_1})f(x,v)f(x,v_1)\;.
 \end{equation}
By a straightforward integration by parts, we get that
\begin{equation}
\label{eq:GCL}
 \int dv\, u(v)\,Q_B^\varepsilon (f,f)(v)  \approx  \int dv\, u(v)\,Q_L (f,f)(v)\;.
\end{equation}
The collision operator $Q_L$ has been introduced by Landau in 1936
for the study of a weakly interacting dense plasma \cite {LL} and 
$$
(\partial_t+v\cdot \nabla_x)f=Q_L(f,f)
$$
 is called the {\em Landau equation} (sometimes, Fokker-Planck-Landau equation)\footnote{The Landau equation was obtained from the Boltzmann equation for cutoffed Coulomb potential (truncated both at short and large distances). Actually the word ``Coulomb'' is frequently used for the Landau equation with kernel singularity $\frac 1 {|U|}$ (see \eqref{a}), which is somehow misleading. In fact as we have seen, this singularity is always present.}.

The qualitative properties of the solutions to the Landau equation are the
same as for the Boltzmann equation regarding the basic conservation
laws and the $H$ theorem.

The procedure described above is a {\em grazing collision} limit. To the best of our knowledge, there is no rigorous version of the formal statement \eqref{eq:GCL}. The available rigorous results on grazing collision limits concern a suitable rescaling of the differential cross-section (rather than the potential): see \cite {LBH} and references therein. 

Even a rigorous proof of  \eqref{eq:GCL} would be not completely satisfactory. Indeed the Landau equation is expected to be a fundamental equation, derivable from particle systems in the weak-coupling limit. A rigorous proof of this fact seems to be hard, even for short times. We will present a formal derivation, outlining the difficulties, in Section \ref{subsec:wclcs}.
%
%
%
%
%
%
%
%

\subsection{Remarks on the scaling limits}

%
%

Let us give a unified picture of the different regimes discussed so  far, leading to  the Boltzmann and the Landau equation. 

The starting point is always a classical system of $N$ identical particles of unit
mass. Microscopic positions and velocities are denoted by $q_1,\cdots, q_N$ and
$v_1,\cdots, v_N$. Let $\tau$ be the microscopic time. The Newton's equations read:
\begin{equation}
\label{newton}
\frac {d}{d\tau} q_i=v_i\;, \qquad \frac {d}{d\tau} v_i=\sum _{\substack
{j=1, \cdots, N \\ j\neq i}} F(q_i-q_j)
 \end{equation}
 where $F=-\nabla \phi$ denotes the interparticle (conservative) force,
$\phi$ the two-body, spherically symmetric potential. 

There is a unique scaling parameter  $ \varepsilon$, which can be interpreted as the ratio between typical macroscopic and microscopic units. In practice we introduce macroscopic variables 
$$
x= \varepsilon \,q,   \quad  t=  \varepsilon\, \tau\;,
$$
and $ \varepsilon$ has to be sent to zero to extract the essential macroscopic features.
Note that the velocity remains unscaled.
In these new variables, the system reads:
\begin{equation}
\label{newtonres}
\frac {d}{dt} x_i=v_i\;, \qquad \frac {d}{dt } v_i= \frac 1  \varepsilon\sum _{\substack
{j=1, \cdots, N \\ j\neq i}}  F\left(\frac {x_i-x_j}  \varepsilon\right)\;. 
 \end{equation}

In order to have a finite  density we should postulate $N \sim \varepsilon^{-3}$ in three dimensions.
Instead, in the low-density limit we chose $N \sim \varepsilon^{-2}$ so that,
for a test particle, the change of momentum (or velocity) for each collision is 
$
\delta v \sim  \frac 1  \varepsilon \delta t=O(1)\;,
$
where the typical interaction time $\delta t$ is $O( \varepsilon)$ (if $F$ has short range); 
on the other hand the collision frequency scales as the number of particles
in the tube of radius $\varepsilon$ (which has volume $ \varepsilon^2$), therefore it is finite in the Boltzmann-Grad limit.
We have been dealing with this scaling in the most favourable situation, the system of hard spheres. In this case, the collision is instantaneous with transferred momentum of $O(1)$.

%

We are now interested in a situation where the interaction is very weak
for which we rescale the potential as 
$
\phi \to  \varepsilon ^\alpha \phi\;,
$ $\alpha \in (0,1)$
and the equations of motion become
\begin{equation}
\label{newtonres1}
\frac {d}{dt} x_i=v_i\;, \qquad \frac {d}{dt} v_i= \varepsilon^{\alpha-1}\sum _{\substack
{j=1, \cdots, N \\ j\neq i}}  F\left(\frac {x_i-x_j}  \varepsilon\right)\;.
 \end{equation}
 We should scale the number density as $N^{-\beta}$ with suitable $\beta$, to get a kinetic equation.
The heuristic argument for the weak-coupling limit discussed in the previous section implies that,
setting $\beta=2(1+\alpha)$, one should get diffusion in velocity, preserving mass, momentum and energy.
Thus we expect that this regime is ruled out by the Landau equation, with the only exception $\alpha=0$, for which we recover 
the low-density scaling and the Boltzmann equation. Frequently, ``weak-coupling limit'' refers to the special case
$\alpha= 1/2$, which is also the case considered in the next section.

\subsection{Weak-coupling limit for classical systems} \label{subsec:wclcs}

We start from the weak-coupling dynamics in macroscopic variables
\begin{equation}
\label{WCo}
\frac {d}{dt} x_i=v_i\;, \qquad \frac {d}{dt} v_i=-\frac 1{\sqrt {\varepsilon}}\sum _{\substack
{j=1, \cdots, N \\ j\neq i}}  \nabla \phi\left(\frac
{x_i-x_j}{\varepsilon}\right)\;, 
 \end{equation}
where we pose $N=\varepsilon^{-3} $.

Once again, $W^N=W^N (Z_N)$ is a symmetric probability density on the phase space $\mathbb{R}^{3N}\times \mathbb{R}^{3N}$,
obeying the Liouville equation

\begin{equation}
\label{Liou}
\left(\partial_t+\sum_{i=1}^N v_i \cdot \nabla_{x_i}\right) W^N=\frac 1{\sqrt\varepsilon} \,
T^\varepsilon_NW^N
 \end{equation}
where
\begin{equation}
T^\varepsilon_NW^N=\sum_{1 \leq k< \ell\leq N}
T^\varepsilon_{k, \ell}W^N\;,
 \end{equation}
\begin{equation}
T^\varepsilon_{k, \ell}W^N=\nabla \phi\left(\frac {x_k-x_ \ell}{\varepsilon}\right)
\cdot (\nabla_{v_k}- \nabla_{v_ \ell})W^N . 
 \end{equation}
The marginals $\left(f_j^N (t)\right)_{j=1}^N $ satisfy the BBGKY hierarchy 
\begin{equation}
\label{eq:bbgkywc}
\left(\partial_t+\sum_{k=1}^j v_k\cdot \nabla_k\right)f^N_j=
\frac {1}{\sqrt {\varepsilon}} T^{\varepsilon}_j f^N_j + \frac {N-j}{\sqrt
{\varepsilon}}\varepsilon^3\,C^{\varepsilon}_{j+1}f^N_{j+1}\;,
 \end{equation}
where the operator $C^{\varepsilon}_{j+1}$ is now defined by
$$C^{\varepsilon}_{j+1}=\sum_{k=1}^j C^{\varepsilon}_{k,j+1}
\; ,
$$ 
\begin{align}
C^{\varepsilon}_{k,j+1}f^N_{j+1}(Z_j)&=-\varepsilon^{-3}\iint dv_{j+1} dx_{j+1}
F \left(\frac {x_k-x_{j+1}}{\varepsilon}\right)
\cdot \nabla_{v_k}
f^N_{j+1}(Z_{j},x_{j+1},v_{j+1})
\nonumber \\
& = - \iint dv_{j+1} dX\,F (X)
\cdot \nabla_{v_k}
f^N_{j+1}(Z_j,x_k - \varepsilon X, v_{j+1})\;.
 \end{align}

One should note that this hierarchy has the same structure of \eqref{hie2}, but now we are considering a smooth and weakly rescaled potential $\phi$.
 In fact $C^\varepsilon_{k,j+1}$ describes the
``collision'' of particle $k$, belonging to the $j$-particle
subsystem, with a particle outside the subsystem, conventionally
$j+1$. The dynamics of the $j$-particle
subsystem is governed by three effects: the free-streaming operator, the
collisions ``inside'' the subsystem (the $T^\varepsilon_j$ term), and the
collisions with particles ``outside'' the subsystem (the $C^\varepsilon_{j+1}$
term).

We can complement the above equations with the initial condition 
\begin{equation}
 f^N_{ j}(t=0) =f_0^{\otimes j}\;,
 \end{equation}
where $f_0$ is a given one-particle density. 
Particles are statistically uncorrelated at time zero, and statistical independence breaks at time $t>0$ because of the dynamics. Since the interaction between two given particles is vanishing in the limit $\varepsilon \to 0$, we can hope for propagation of chaos. The 
physical mechanism producing chaos is however quite different from the one discussed in Section \ref{sec:2}.
Here, two given particles can interact, the force is strong but the net effect of the collision is small (because the interaction time is small), while in the low-density regime collisions are always strong and unlikely.


Let us investigate the convergence of $f^N_1$  to the Landau equation, in the limit $\varepsilon \to 0$, using the hierarchy \eqref{eq:bbgkywc}.

Expanding $f^N_j(t)$ as a perturbation of the free flow $S(t)$ (as in \eqref{seriesB})
we find that
\begin{align}
\label{WCcoll}
f_j^N(t)= &S(t)f_0^{\otimes j}+ \frac {N-j}{\sqrt {\varepsilon}} \varepsilon^3 \int _0^t
S(t-t_1)C_{j+1}^\varepsilon f_{j+1}^N (t_1)dt_1+\\
&\frac {1}{\sqrt {\varepsilon}}\int _0^t S(t-t_1)T_{j}^\varepsilon f_{j}^N (t_1)dt_1\nonumber\;.
 \end{align}
It is now reasonable to assume that
\begin{equation}
\int dX \,F(X)=0\;,\nonumber
 \end{equation}
which implies
$
C_{j+1}^\varepsilon f_{j+1}^N=O(\varepsilon)\;,
$
{\em provided} that the second derivatives $D^2_v f_{j+1}^N(t)$ are bounded uniformly in $\varepsilon$. 
Since 
$
\frac {N-j}{\sqrt {\varepsilon}}=O(\varepsilon^{- \frac 72})\;,
$
we see that the second term in the right-hand side of  \eqref {WCcoll}  does not give
any contribution in the limit.
In the same assumptions,
\begin{align}
&\int _0^t S(t-t_1)T_{j}^\varepsilon f_{j}^N (t_1)dt_1=\nonumber\\
&\sum_{i \neq k}\int _0^t
F \left(\frac {(x_i-x_k)-(v_i-v_k) (t-t_1)}{\varepsilon} \right)
\cdot g(Z_j,t_1)dt_1 \nonumber
 \end{align}
where $g$ is a smooth $j$-particle function, which is again $O(\varepsilon)$ so that the last
term in the right-hand side of   \eqref {WCcoll}  is also vanishing in
the limit. We are therefore facing the alternative: either the limit is
trivial, or the time evolved marginals are not smooth.
This is indeed bad news: a rigorous derivation of the (expected) Landau equation seems problematic.

The above difficulty suggests to split $f^N_j(t)$ into two parts,
namely we conjecture that:
\begin{equation}
f_j^N=g_j^N+\gamma_j^N\;,  \nonumber
 \end{equation}
where $g_j^N$ is the main part of $ f_j^N$ and is smooth, while $\gamma_j^N$ is small, but strongly oscillating (hence with large derivatives). 
The two parts satisfy, by definition,
\begin{equation}
\label{eqsplit}
\left(\partial_t+\sum_{k=1}^j v_k\cdot \nabla_{x_k}\right)g^N_j=
 \frac {N-j}{\sqrt {\varepsilon}}\varepsilon^3 C^{\varepsilon}_{j+1}g^N_{j+1}+ 
  \frac {N-j}{\sqrt {\varepsilon}}\varepsilon^3 C^{\varepsilon}_{j+1}\gamma^N_{j+1}\;, \nonumber
 \end{equation}
\begin{equation}
\left(\partial_t+\sum_{k=1}^j v_k\cdot \nabla_{x_k}\right)\gamma^N_j=
\frac {1}{\sqrt {\varepsilon}} T^{\varepsilon}_j \gamma^N_j + \frac {1}{\sqrt {\varepsilon}} T^{\varepsilon}_j g^N_j \;, \nonumber
 \end{equation}
with initial data
\begin{equation}
g^N_j =f_0^{\otimes j}, \quad \gamma^N_j =0 \nonumber \;.
 \end{equation}
The remarkable feature of this decomposition is that the singular part can be eliminated. In fact we have that
\begin{align}
f_1^N(t)= &S(t)f_0+ \frac {N-1}{\sqrt {\varepsilon}}\varepsilon^3 \int _0^t
S(t-t_1)\,C_{2}^\varepsilon \left(g_{2}^N (t_1)+\gamma^N _2 (t_1)\right) dt_1  \nonumber\;,
 \end{align}
where
$$
\gamma^N_2(t)=\frac 1 {\sqrt {\varepsilon}} \int_0^t ds 
\,U^\varepsilon_2(s)\,T^ \varepsilon_2 g_2^N(t-s)
$$
and $U^\varepsilon_2$ is just the two-particle interacting flow. Indicating by $\left(Z_2^\varepsilon(-s)\right)_{s \in (0,t)}$
this flow with final condition $Z_2^\varepsilon(0) = Z_2$, we have that 
\begin{align}
\label{gamma}
\gamma^N_2 (Z_2 ,t)& =\frac 1 {\sqrt {\varepsilon}} \int_0^t ds 
\,\nabla \phi\left(\frac
{x_1^\varepsilon(-s) -x^\varepsilon_2(-s) }{\varepsilon}\right) \cdot \left(\nabla_{v_1}-\nabla_{v_2} \right)\,g_2^N(Z_2^\varepsilon(-s), t-s) \nonumber.
 \end{align}
%

Based on the conjecture, we present now a formal derivation of the Landau equation (assuming $g_2^N $smooth).
We have that
\begin{align}
\left(\partial_t+v_1\cdot \nabla_{x_1}\right)f^N_1 (t)=
 \frac {N-1}{\sqrt {\varepsilon}}\varepsilon^3 C^{\varepsilon}_{2}g^N_{2} (t)+ \frac {N-1}{\varepsilon} \varepsilon^3 C^ \varepsilon_2  \int_0^t ds  \,U^\varepsilon_2 (s)\,T^\varepsilon_2 g^N_2 ( t-s) \nonumber \;.
 \end{align}
Let $u\in {\cal D}$ be a test function. As already mentioned the first term on the right-hand side is negligible:
$$
\frac {N-1}{\sqrt{\varepsilon}}\varepsilon^3\left(u,C^{\varepsilon}_{2}g^N_{2} (t) \right)=O\left(\sqrt { \varepsilon}\right)\;.
$$
The last term gives
\begin{eqnarray}
&&-\frac {N-1}{\varepsilon} \int dz_1 \int dz_2 \int_0^t ds \, \nabla_{v_1} u(z_1) \nonumber\\
&&\ \ \ \ F \left(\frac {x_1 -x_2 }{\varepsilon}\right)  F \left(\frac
{x_1(-s) -x_2(-s) }{\varepsilon}\right) \cdot (\nabla_{v_1}-\nabla_{v_2} )g_2^N(Z^\varepsilon_2(-s),t-s) \nonumber\\
&&\approx -\int dz_1  \int dr \,dv_2   \int_0^\infty ds \,\nabla_{v_1} u(z_1)  \nonumber\\
&& \ \ \ \ F (r)\,  F \left(\frac{x^\varepsilon_1(- \varepsilon s) -x^\varepsilon_2(- \varepsilon s) }{\varepsilon}\right) \cdot (\nabla_{v_1}-\nabla_{v_2} )g_2^N(x_1,v_1,x_1,v_2,t)\;,
\nonumber
\end{eqnarray}
after having changed to variables $r=\frac {x_1-x_2}{ \varepsilon}$ and $s \to \frac s  \varepsilon$. Here,
setting $U=v_1-v_2$,
$$
\frac {x^\varepsilon_1(- \varepsilon s) -x^\varepsilon_2(- \varepsilon s) }{\varepsilon} \approx r+Us \;.
$$
This term is then approximately equal to
\begin{eqnarray}
&&  -\int dz_1  \int dr dv_2   \int_0^\infty ds\nabla_{v_1} u(z_1)
 F (r)  F \left(r + Us\right) \cdot (\nabla_{v_1}-\nabla_{v_2} )g_2^N(x_1,v_1,x_1,v_2,t)\nonumber \\
&&\approx \left(u,Q_L(g_1^N,g_1^N)\right)\nonumber\;,
 \end{eqnarray}
where in the last step we invoked propagation of chaos ($ g_2^N \approx (g_1^N )^{\otimes 2}$)
and used definition \eqref{eq:LO}.
Indeed it is not hard to show that
$$
\int dr  \int_0^{\infty} ds F( r) F( r+ Us)=
\frac 12  \int dr  \int_{-\infty}^{\infty} ds F( r) F( r + Us)= a(U)
$$
where $a(U)$ is the matrix given by \eqref{a}. Indeed expressing the above identity in terms of the Fourier 
transforms we readily arrive to the right-hand side of \eqref {T}. 

Unfortunately, very little is known about the mathematical derivation. We mention here 
the only result we are aware of.

Consider the first order (in time) approximation $\tilde g^N_j$ of  $g^N_j$ given by
\begin{eqnarray}
&&\tilde g^N_j(t)=S(t)f_0^{\otimes j} + 
\frac{N-j}{\sqrt\varepsilon}\varepsilon^3 \int_0^t S(t-\tau) C^\varepsilon_{j+1} S(\tau)g^N_{j+1}\,d\tau\nonumber\\
 &&\ \ \ \ +\frac{N-j}{\varepsilon}\varepsilon^3 \int_0^t d\tau \int_0^\tau d\sigma
 S(t-\tau) C^\varepsilon_{j+1}U^\varepsilon_{j+1}(\tau-\sigma)T^\varepsilon_{j+1}S(\sigma)f_0^{\otimes (j+1)}\;.
 \end{eqnarray}
Then we can prove:

\smallskip
{\bf Theorem \cite{BPS}. }
{ \it Suppose that 
$f_0\in C^3_0(\mathbb{R}^3 \times \mathbb{R}^3)$ is the initial probability density
satisfying:
\begin{equation}\label{phi-property}
| D^r f_0 (x,v)| \leq C e^{-b |v|^2} \quad \text {for} \qquad  r=0,1, 2
 \end{equation}
where $D^r$ is any derivative of order $r$ and $b >0 $. Assume
 $\phi \in C^2(\mathbb{R}^3)$
and $\phi(x)=0$ if $|x| >1$.
Assume that the marginals factorize exactly at time zero.
Then
\begin{equation}
\label{result}
\lim_{\varepsilon \to 0} \tilde g^N_1(t)=S(t) f_0+\int_0^t d\tau S(t-\tau) Q_L( S(\tau)f_0, S(\tau)f_0)
 \end{equation}
where $N\varepsilon^3=1$ and the above limit is considered in ${\cal D}'$.
}

\smallskip

Since the right-hand side of Eq.\,\eqref{result} is the first order approximation of the
Landau equation, we can consider the theorem as a consistency result. 

\end{document}